\documentclass[a4paper, amsfonts, amssymb, amsmath, reprint, showkeys, footinbib, twoside,superscriptaddress,floatfix,longbibliography]{revtex4-1}
\usepackage{graphicx}%
\usepackage{dcolumn}%
\usepackage{bm}%
\usepackage{float}
\usepackage{xcolor}
\usepackage{mhchem}
\usepackage{algorithmic}
\usepackage{gensymb}
\usepackage{booktabs}
\usepackage{verbatim}
\usepackage{newtxtext}
\usepackage[slantedGreek]{newtxmath}
\usepackage{amsmath}
\usepackage{longtable}
\usepackage[utf8]{inputenc}

\usepackage[T1]{fontenc}
\usepackage{hyphenat}
\usepackage{comment}
\usepackage[pdftex, bookmarks, pdffitwindow=false, pdfstartview=FitH, pdfdisplaydoctitle, colorlinks, plainpages=false, pdftitle={},pdfauthor={}, pdfpagelabels, hypertexnames, citecolor={blue!50!black},linkcolor={blue!50!black}, urlcolor={blue!50!black}, pdflang={en}, hyperfootnotes=false, breaklinks]{hyperref}

\usepackage{siunitx}

\DeclareUnicodeCharacter{0308}{HERE!HERE!} 
\DeclareUnicodeCharacter{2212}{{\ensuremath{-}}}
\DeclareUnicodeCharacter{0308}{{\ensuremath{~}}}

\newcommand{\Rmnum}[1]{\expandafter\@slowromancap\romannumeral #1@}

\DeclareSIUnit\angstrom{\text {Å}}

\makeatletter
\renewcommand{\@seccntformat}[1]{}
\makeatother

\makeatletter
\def\@bibdataout@aps{
 \immediate\write\@bibdataout{
 @CONTROL{
   apsrev41Control, author="48",editor="1",pages="0",title="0",year="1"
 }}
 \if@filesw
  \immediate\write\@auxout{\string\citation{apsrev41Control}}
 \fi
}
\makeatother

\begin{document}

\preprint{}

\title{Strong Crystalline Thermal Insulation Induced by Extended Antibonding States}

\author{Ruihuan Cheng}
\affiliation{Department of Mechanical Engineering, The University of Hong Kong, Pokfulam Road, Hong Kong SAR, China}
\author{Chen Wang}
\affiliation{Department of Mechanical Engineering, The University of Hong Kong, Pokfulam Road, Hong Kong SAR, China}
\affiliation{Institute for Advanced Study, Shenzhen University, Shenzhen 518060, China}
\author{Niuchang Ouyang}
\affiliation{Mechanical Engineering and Materials Science, Duke University, Durham, North Carolina, USA}
\author{Xingchen Shen}
\email{xingchen.shen@nwpu.edu.cn}
\affiliation{MOE Key Laboratory of Material Physics and Chemistry under Extraordinary Conditions, School of Physical Science and Technology, Northwestern Polytechnical University, Xi’an 710072, China}
\author{Yue Chen}
\email{yuechen@hku.hk}
\affiliation{Department of Mechanical Engineering, The University of Hong Kong, Pokfulam Road, Hong Kong SAR, China}

\date{\today}

\begin{abstract}

 Crystalline solids with extreme insulation often exhibit a plateau or even an upward-sloping tail in thermal conductivity above room temperature. Herein, we synthesized a crystalline material AgTl$_2$I$_3$ with an exceptionally low thermal conductivity of 0.21 $\rm W m^{-1} K^{-1}$ at 300 K, which continues to decrease to 0.17 $\rm W m^{-1} K^{-1}$ at 523 K. We adopted an integrated experimental and theoretical approach to reveal the lattice dynamics and thermal transport properties of AgTl$_2$I$_3$. Our results suggest that the Ag-I polyhedron enables extended antibonding states to weaken the chemical bonding, fostering strong lattice anharmonicity driven by the rattling vibrations of Ag atoms and causing lattice softening. Experimental measurements further corroborate the large atomic thermal motions and low sound velocity. These features impede particle-like phonon propagation, and significantly diminish the contribution of wave-like phonon tunneling. This work highlights a strategy for designing thermal insulating materials by leveraging crystal structure and chemical bonding, providing a pathway for advancing the development of thermal insulators.

\end{abstract}

\maketitle
\section{Introduction}
The quest for crystalline solids with ultralow thermal conductivity ($\kappa$) is crucial for applications in thermoelectrics \cite{zhao2014ultralow, aenm201701797, yan2022high}, thermal insulation \cite{scienceabh1619, PRL130236301}, and energy-efficient technologies \cite{science1068609, CLARKE200522}. Achieving ultralow $\kappa$ requires a thorough understanding of the underlying mechanisms of thermal transport at the atomic level. In the kinetic theory \cite{tritt2005thermal}, insulating materials with low phonon group velocity, wide phonon linewidth, and small heat capacity usually have low $\kappa$. Following this principle, several strategies have been employed to suppress thermal transport, including complex crystal structures \cite{adfm202108532, aenm201800030}, anharmonic lattice dynamics \cite{PRL125045701, PRL125245901}, and atomic rattling effects \cite{jacs1c11887, advs202406380}.

Complex materials with large unit cells, weak chemical bonding, heavy elements, and rattling phonon modes, such as clathrates \cite{PRL82779, CHAKOUMAKOS200080}, chalcogenides \cite{anie201605015, jacs3c04871}, tetrahedrites \cite{PRL125085901, D4TA03316G}, and skutterudites \cite{science27252661325, D2TA02687B} are good thermal insulators. For instance, ultralow thermal conductivity of ~0.2 $\rm W m^{-1} K^{-1}$ is witnessed in the complex crystalline materials CsAg$_5$Te$_3$ \cite{anie201605015, nsrnwae216} and Cs$_3$Bi$_2$I$_9$ \cite{adfm202304607} at room temperature. However, the pursuit of structural complexity inevitably introduces significant coherence contribution to $\kappa$, arising from wave-like tunneling of densely packed adjacent phonons, as suggested by the unified thermal transport theory proposed by Simoncelli et al. \cite{UT2019} and Isaeva et al. \cite{isaeva2019modeling}. This leads to a notable enhancement in the thermal transport, with plateaus observed above room temperature in the thermal conductivities of CsAg$_5$Te$_3$ and Cs$_3$Bi$_2$I$_9$, or even an increasing trend in argyrodites \cite{petrov1975characteristics, advs202400258, PRB111064307}. 

Alternatively, searching for strong anharmonic effect in simple crystalline solids is another way to achieve ultralow-$\kappa$ materials. For example, the primitive cells of CuI \cite{perry2016handbook}, InTe \cite{zhang2021direct}, AgCrSe$_2$ \cite{cm200581k}, GeSe \cite{jacs0c03696}, TlSe \cite{jacs9b10551}, AgI \cite{GOETZ1982293}, Tl$_3$VSe$_4$  \cite{scienceaar8072}, and Cs$_2$SnI$_6$ \cite{acschemmater2c00084} contain no more than 10 atoms, and they exhibit ultralow $\kappa$ at room temperature. These materials have strong anharmonic phonon vibration modes caused by rattling atoms \cite{PRL114095501, PRL117046602, jacs1c11887}, long-range pair interactions \cite{PRL107235901, anie201511737, zhang2023dynamic}, ferroelectric instability-induced phonon softening \cite{delaire2011giant, jacs0c03696}, and metastable intrinsic defect geometries \cite{PRL130236301}, which effectively suppress particle-like phonon propagation. Due to their simple crystal structures, the wave-like tunneling effects in these materials are relatively weak. Our recent work reported that AgTlI$_2$ with a simple crystal structure has an ultralow thermal conductivity of 0.25 $\rm W m^{-1} K^{-1}$ at 300 K \cite{zeng2024pushing}. 

These insights motivate a structure-informed approach to the discovery of thermal insulators. While previous efforts have predominantly focused on lattice complexity and strong anharmonic effects, an implicit and often underappreciated factor is the presence of weak chemical bonding \cite{SPITZER197019, adfm202108532}. Weak bonding typically results in reduced phonon group velocities, which inherently suppress the particle-like phonon propagation. Moreover, low phonon group velocities lead to the reduced velocity matrix elements of coherent modes, thereby diminishing the contribution of wave-like tunneling to thermal transport. For instance, materials such as AgI \cite{GOETZ1982293}, Cs$_3$Bi$_2$I$_9$ \cite{adfm202304607}, and Cs$_2$SnI$_6$ \cite{acschemmater2c00084}, characterized by soft lattice and low elastic moduli, display both strong anharmonic phonon behavior and intrinsically low sound velocities, resulting in ultralow lattice thermal conductivities. This dual suppression of particle-like and wave-like phonon transports underscores the necessity of considering weak bonding as a synergistic strategy alongside anharmonicity for discovering high-performance thermal insulators.

\begin{figure*}[htbp]
\includegraphics[width=0.8\linewidth]{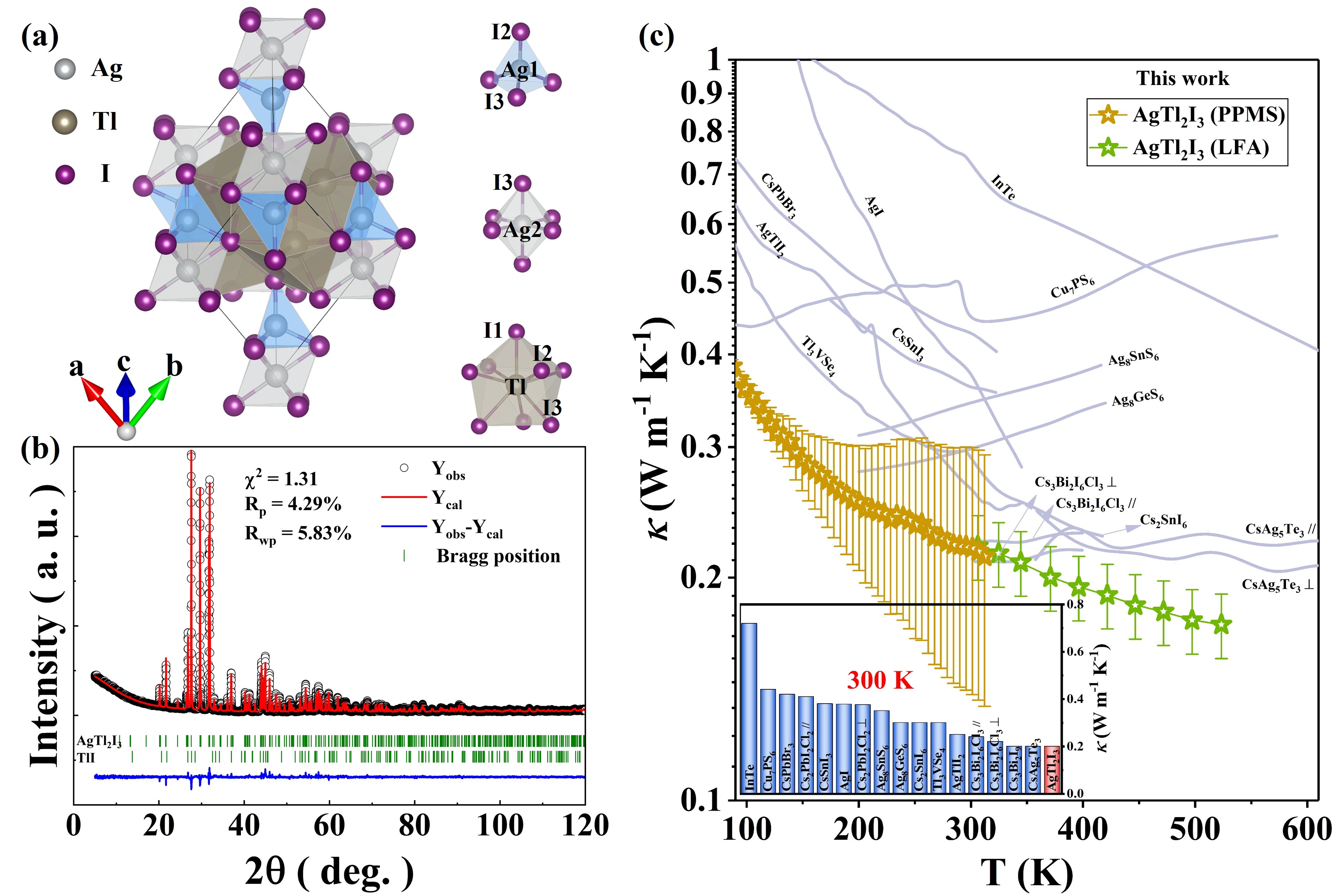}
\caption{\label{fig:1} \textbf{Crystal structure and thermal conductivity of AgTl$_2$I$_3$.} (a) Crystal structure of AgTl$_2$I$_3$ visualized using VESTA \cite{VESTA} and distinct coordination environments of Ag1, Ag2, and Tl atoms. (b) Rietveld refinements of the AgTl$_2$I$_3$ powder sample at 300 K. (c) Temperature-dependent thermal conductivity $\kappa$ of AgTl$_2$I$_3$ compared to those of dense thermal insulators \cite{zhang2023dynamic, PRB107064308, advs202400258, pnas1711744114, zeng2024pushing, scienceaar8072, petrov1975characteristics, acharyya2022glassy, acschemmater2c00084, nsrnwae216, anie201605015, jacs0c08044, adfm202304607}. The inset shows the values at 300 K.}
\end{figure*}

Herein, we synthesized a novel crystalline solid AgTl$_2$I$_3$ that exhibits an ultralow $\kappa$ of 0.21 $\rm W m^{-1} K^{-1}$ at 300 K, which further decreases to 0.17 $\rm W m^{-1} K^{-1}$ at 523 K. To reveal the microscopic mechanisms of the suppressed thermal transport in AgTl$_2$I$_3$, we performed experimental measurements, first-principles anharmonic lattice dynamics calculations, molecular dynamics simulations based on a machine learning neuroevolution potential (NEP) \cite{fan2021NEP2}, and chemical bonding analysis. By incorporating temperature renormalization effects, our study successfully rationalizes the experimentally observed ultralow thermal conductivity of AgTl$_2$I$_3$, offering a robust theoretical basis for understanding the microscopic origin of its thermal insulating properties. Structural and chemical bond analysis reveals that the face-sharing Ag-I polyhedral crystal structure enables extended antibonding states, leading to weak bonding and strong lattice anharmonicity. Theoretical anharmonicity parameters and experimentally measured atomic thermal vibrations indicate that the rattling motion of Ag atoms dominates the thermal insulation. Additionally, this unique crystal structure and weak bonding result in an exceptionally low phonon group velocity, which is experimentally confirmed by further measurement of the sound velocity. The face-shared Ag-I polyhedral crystal structure and weakened chemical bonding give rise to strong anharmonicity and low phonon group velocity, thereby suppressing both particle-like propagation and wave-like tunneling in phonon transport.

\section{Results and Discussion}
\subsection{Crystal Structure and Thermal Conductivity}

\begin{figure*}[htbp]
\includegraphics[width=1.0\linewidth]{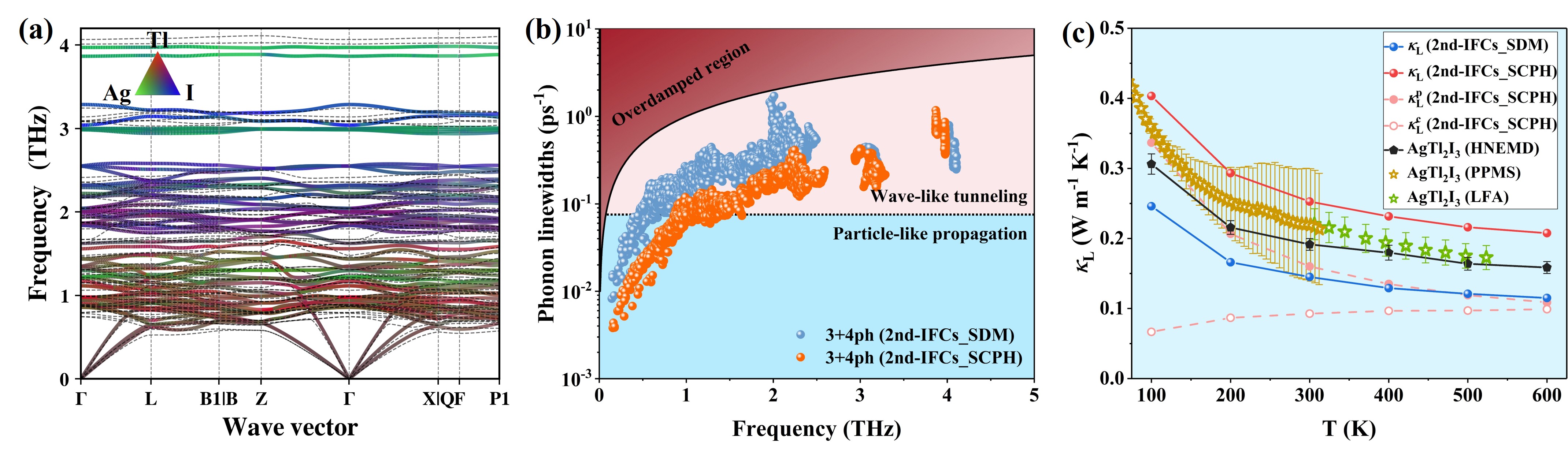}
\caption{\label{fig:2} \textbf{Anharmonic phonon renormalization and dual-channel thermal transport in AgTl$_2$I$_3$.} (a) Phonon dispersions of AgTl$_2$I$_3$ at 300 K calculated using the SCPH method. The harmonic phonon dispersions at 0 K (black dotted lines) calculated by the small displacement method (SDM) are plotted for comparison. (b) Phonon linewidths (including three- and four-phonon scatterings) of AgTl$_2$I$_3$ at 300 K calculated using perturbation theory (PT) and phonon frequencies obtained via SDM and SCPH schemes. The solid line represents the phonon linewidth equal to the corresponding phonon frequency. The dashed line represents the Wigner limit defined as the average interband spacing \cite{PRX12041011}. (c) Average lattice thermal conductivity of AgTl$_2$I$_3$ calculated using the unified thermal transport theory with different sets of phonon properties, alongside the results obtained from homogeneous nonequilibrium molecular dynamics (HNEMD) simulations. Our experimental thermal conductivity data are also shown for comparison.}
\end{figure*}

AgTl$_2$I$_3$ crystallizes in the $R\overline{3}$ space group with a unit cell containing 18 atoms (Fig. \ref{fig:1}a). Rietveld refinement of the high-resolution powder X-ray diffraction (PXRD) pattern confirms that the synthesized polycrystalline sample (Fig. \ref{fig:1}b) adopts the $R\overline{3}$ phase and verifies that the sample purity is high, with only trace amounts of the TlI impurities (less than 1\%). AgTl$_2$I$_3$ can be described as a framework composed of anionic [Ag$_3$I$_8$]$^{5-}$ and cationic [Tl$_6$I]$^{5+}$ units. In the anionic [Ag$_3$I$_8$]$^{5-}$, two Ag$^+$ ions (Ag1) adopt a tetrahedral geometry (Ag1-I), which interconnect via face-sharing with a third Ag$^+$ (Ag2) residing in an octahedron environment (Ag2-I). Meanwhile, the Tl$^+$ cations are coordinated by eight nearest I$^-$ neighbors at distances ranging from 3.39 to 3.78 \AA, forming a distorted Tl-I octahedron. The Ag-Ag distance (3.00 \AA) is notably longer than that in metallic Ag (2.88 \AA) \cite{ic00339a009, wells2012structural}, suggesting weak Ag-Ag interactions. The Coulombic repulsion between Ag1 and Ag2 causes an elongation of the Ag-I bonds \cite{ja01379a006}, resulting in two inequivalent bond lengths (2.75 and 2.84 \AA) of Ag1-I. Meanwhile, the octahedron Ag2 center forms a uniform and longer bond (3.08 \AA) with I atoms due to the relatively high coordination number. The polyhedral crystal structure of the cationic [Ag$_3$I$_8$]$^{5-}$ unit creates a flexible structural environment \cite{SPITZER197019, anie201508381}, facilitating rattling-like thermal vibration of Ag atoms. This is supported by the large experimental atomic displacement parameter of Ag atoms from Rietveld refinements (Table S1). The underlying weak bonding characteristics play a crucial role in determining the lattice dynamics and thermal transport properties of AgTl$_2$I$_3$.

The thermal conductivity of the sintered polycrystal sample AgTl$_2$I$_3$, ranging from 100 to 523 K, is shown in Fig. \ref{fig:1}c. AgTl$_2$I$_3$ exhibits an ultralow thermal conductivity of 0.21 $\rm W m^{-1} K^{-1}$ at room temperature, which is lower than that of AgTlI$_2$ \cite{zeng2024pushing} and Cs$_3$Bi$_2$I$_6$Cl$_3$ \cite{acharyya2022glassy}. Its thermal conductivity is comparable to the values observed in CsAg$_5$Te$_3$ \cite{anie201605015, nsrnwae216} and Cs$_3$Bi$_2$I$_9$ \cite{adfm202304607}, positioning it among the materials with the lowest thermal conductivities at room temperature. Additionally, the thermal conductivity of AgTl$_2$I$_3$ continues to decrease with increasing temperature, reaching an even lower value of  0.17 $\rm W m^{-1} K^{-1}$ at 523 K. This behavior contrasts sharply with materials having complex crystal structures, such as CsAg$_5$Te$_3$, Cs$_3$Bi$_2$I$_9$, and argyrodites \cite{petrov1975characteristics, advs202400258, PRB111064307}, which typically show a plateau or even an increase in thermal conductivity above room temperature due to the wave-like phonon tunneling effect. The continuous decrease in $\kappa$ of AgTl$_2$I$_3$ over a broad temperature range suggests that both phonon particle-like propagation and wave-like tunneling are effectively suppressed.

\subsection{Anharmonic Lattice Dynamics and Dual-Channel Thermal Transport}

To understand the microscopic phonon transport mechanism in AgTl$_2$I$_3$, it is essential to accurately describe its anharmonic lattice dynamics, including the phonon frequency and linewidth, which are particularly important for such a strongly anharmonic material \cite{PRL125045701, PRB105184303, PRB109054305, zeng2024pushing}. To this end, we consider the temperature-dependent phonon frequency renormalization \cite{PRB84180301, PRB92054301, PRB87104111} and the phonon linewidth calculation up to the four-phonon scattering \cite{PRB93045202, PRB96161201}. Figures \ref{fig:2}a and S3 show the renormalized phonon dispersions calculated using the self-consistent phonon (SCPH) theory \cite{PRB1572, SCPH2014} from 100 to 600 K, compared with the harmonic phonon dispersions at 0 K obtained from the small displacement method (SDM). This phonon frequency renormalization alters the phonon scattering phase space, resulting in smaller phonon linewidths, particularly for the low-frequency phonons that undergo substantial hardening, as shown in Fig. \ref{fig:2}b. Significant differences, on the order of magnitude, are observed in the phonon linewidths in the low-frequency region around 1 THz, underlining the critical role of phonon frequency renormalization. Additionally, the frequency-dependent phonon linewidth distribution reveals that the phonon modes exhibit wave-like tunneling starting from approximately 1 THz, as determined by the Wigner criterion \cite{PRX12041011}.

\begin{figure*}[htbp]
\includegraphics[width=0.8\linewidth]{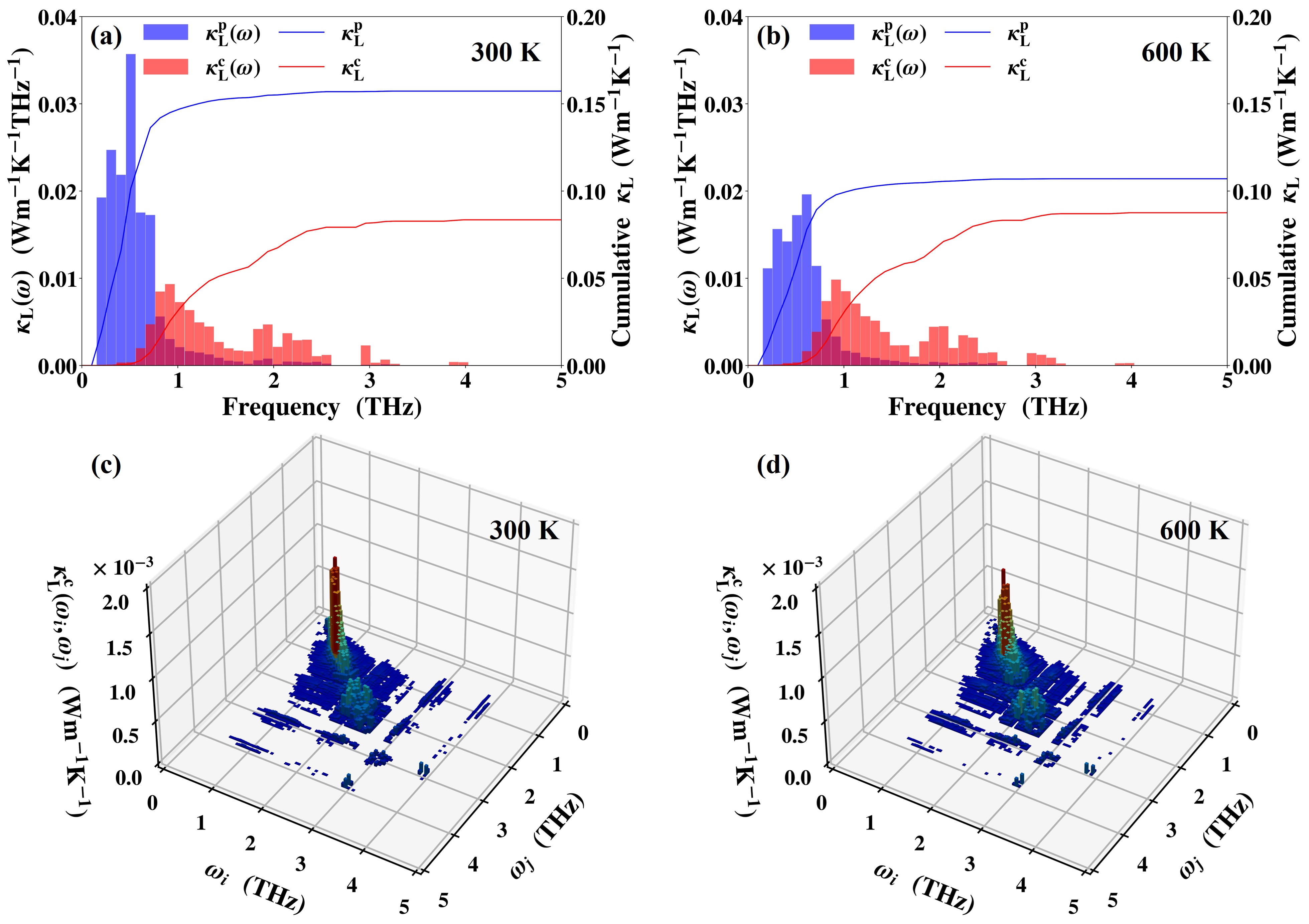}
\caption{\label{fig:3} \textbf{Spectral decomposition and spatial visualization of dual-channel thermal transport in AgTl$_2$I$_3$.} The spectral and cumulative particle-like phonon propagation ($\kappa^{\rm p}_{\rm L}$) and wave-like interband tunneling ($\kappa^{\rm c}_{\rm L}$) contributions to the lattice thermal conductivity along the \textit{x} axis at (a) 300 and (b) 600 K. Three-dimensional visualization model of the contributions to the wave-like interband tunneling thermal conductivity along the \textit{x} axis at (c) 300 and (d) 600 K.}
\end{figure*}

Anharmonic renormalization significantly impacts phonon frequencies and linewidths, which in turn play a critical role in the thermal conductivity. Since AgTl$_2$I$_3$ is an insulator with a large electronic band gap (Fig. S4), heat transports primarily through lattice vibrations; thus, the following discussion focuses on the lattice thermal conductivity ($\kappa_{\rm L}$). We integrated the phonon frequencies (with and without temperature renormalization) and linewidths into the unified thermal transport theory \cite{UT2019}, and calculated the thermal conductivities (Fig. \ref{fig:2}c). Additionally, we have also computed the thermal conductivity of AgTl$_2$I$_3$ using HNEMD simulations based on a machine-learning neuroevolution potential (NEP) (Fig. S5), which accounts for all-order phonon scattering processes. The HNEMD result was found to be lower than the thermal conductivity calculated from the unified thermal transport theory, likely due to the high-order phonon scatterings. Nevertheless, both results show reasonable agreement with the experimental data across a wide temperature range. In contrast, calculations based on harmonic force constants significantly underestimate the experimental lattice thermal conductivity. This result underscores the importance of temperature-dependent phonon properties on the thermal transport of AgTl$_2$I$_3$.

Further insights into the microscopic phonon transport mechanisms are provided by the spectral lattice thermal conductivity, as shown in Figs. \ref{fig:3} and S6. Particle-like propagation is predominantly observed in the low-frequency region, which primarily consists of acoustic phonons with relatively high group velocities. In contrast, wave-like phonon tunneling contributes significantly across the entire phonon frequency range (Fig. \ref{fig:3}a). As the temperature increases from 300 K to 600 K, enhanced phonon scattering leads to significant suppression of particle-like propagation, and the contribution of wave-like phonon tunneling to thermal transport in AgTl$_2$I$_3$ becomes increasingly prominent (Figs. \ref{fig:3}a and b). This shift is driven by the broadening of phonon linewidths, which enhances phonon coherence and facilitates wave-like tunneling processes, particularly in the high-frequency region (> 1.5 THz) (Figs. \ref{fig:3}c and d). However, our analysis shows that particle-like propagation remains the dominant heat transport mechanism across the entire temperature range studied; it is seen that the wave-like tunneling approaches but does not exceed the particle-like contribution at 600 K (Fig. \ref{fig:2}c). This result indicates that wave-like phonon tunneling in AgTl$_2$I$_3$ is less significant despite increased coherence at elevated temperatures, highlighting the overall low efficiency of the phonon transport channels.

\begin{figure*}[htbp]
\includegraphics[width=1.0\linewidth]{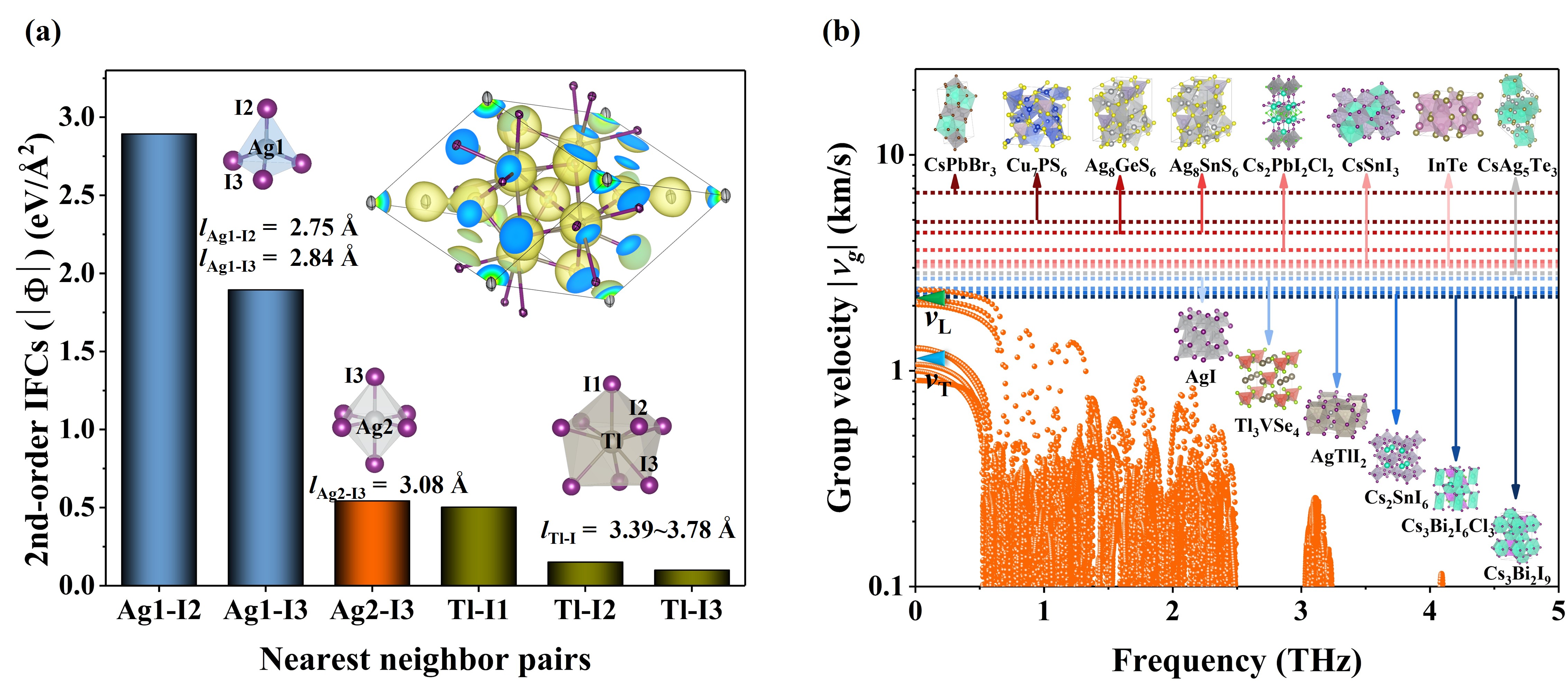}
\caption{\label{fig:4} \textbf{Bonding stiffness, charge distribution, and phonon group velocities of AgTl$_2$I$_3$.} (a) The harmonic interatomic force constants (IFCs) for the nearest neighbor atom pairs in AgTl$_2$I$_3$. The inset shows the total charge density of AgTl$_2$I$_3$, where an isosurface of the charge density has been visualized at 0.051 e/Bohr$^3$. (b) Calculation of the frequency-dependent phonon group velocities of AgTl$_2$I$_3$ and comparison with the maximum of the acoustic phonon group velocities of typical thermal insulation materials at 300 K. The experimentally measured longitudinal ($v_{\rm L}$) and transverse ($v_{\rm T}$) sound velocities of AgTl$_2$I$_3$ are also marked for comparison.}
\end{figure*}

The above results suggest that phonon modes below 1.5 THz dominate thermal transport in AgTl$_2$I$_3$, and a deeper understanding of these modes helps unravel the mystery of its strong thermal insulation. Fig. \ref{fig:2}a shows the phonon dispersion with atomic vibration contribution resolution, indicating that the low-frequency modes in AgTl$_2$I$_3$ are mainly dominated by Tl and Ag atoms.  The tetrahedron and octahedron composed of Ag1 and Ag2 share faces with the polyhedron centered on Tl (Fig. \ref{fig:1}a), leading to a soft lattice \cite{ja01379a006}. Second-order force constants reveal that the strongest interaction in AgTl$_2$I$_3$ occurs between Ag2 and I2 (2.89 eV/\AA$^2$), as indicated by the overlap of charge clouds [Fig. \ref{fig:4}(a)]. However, this force constant is much lower than those of the Bi-I bond of 4.77 eV/\AA$^2$ in Cs$_3$Bi$_2$I$_9$ \cite{adfm202304607} and 5.0 eV/\AA$^2$ in Cs$_3$Bi$_2$I$_6$Cl$_3$ \cite{acharyya2022glassy}. The weak bonding and soft lattice lead to large thermal vibrations of Ag atoms (Table S2). The Tl-I bond is weaker than Ag-I due to the longer bond length, and the thermal vibration of Tl atoms only contributes to the low-frequency phonon modes (Fig. \ref{fig:2}a) and has a much smaller amplitude (Tables S1 and S2), suggesting that Tl atoms act as phonon blockers \cite{PRB106054302}. This weakly bonded structure plays a critical role in suppressing the thermal transport. Since the phonon group velocity (\textit{v}$_{\rm g}$) is proportional to $\sqrt{\textit{k}/\textit{M}}$, where \textit{k} is the bond strength and \textit{M} is the atomic mass \cite{tritt2005thermal}, the weak chemical bonds and large atomic masses are expected to lead to low \textit{v}$_{\rm g}$ in AgTl$_2$I$_3$. Figures S7 and S8, respectively, show the calculated phonon dispersions and frequency-dependent group velocities of typical thermal insulating materials at 300 K. It is found that the highest phonon frequency of AgTl$_2$I$_3$ is lower than that of most of them. Figure \ref{fig:4}b presents the phonon group velocities calculated for AgTl$_2$I$_3$, which are extremely low compared to typical thermal insulating materials, further supporting its observed ultralow thermal conductivity. This is also consistent with the experimentally measured low sound velocities of AgTl$_2$I$_3$, which are 1.138 and 2.169 km/s for the transverse and longitudinal modes, respectively (see Tables S3 and S4).

\subsection{Chemical Bonding Analysis}

\begin{figure*}[htbp]
\includegraphics[width=0.8\linewidth]{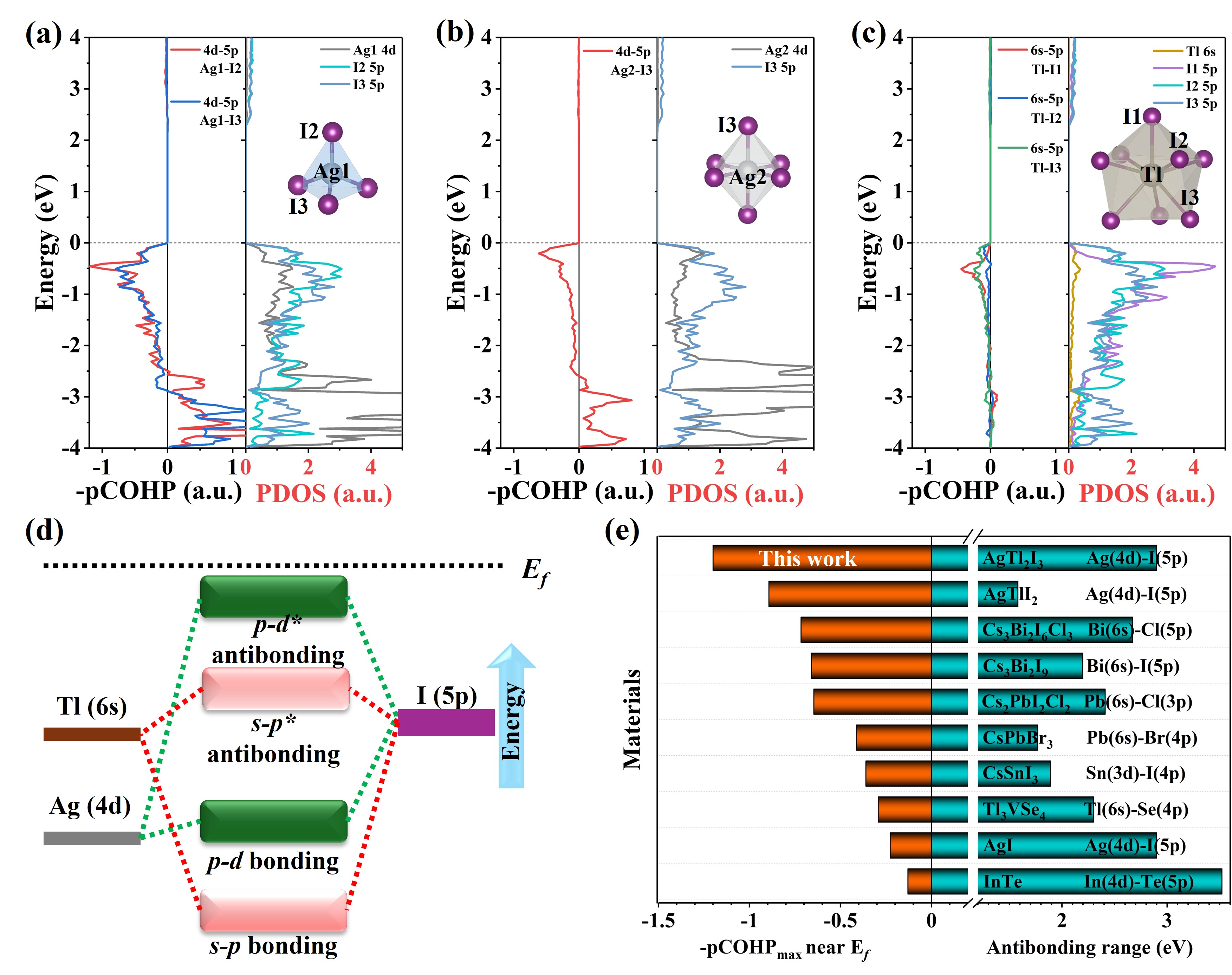}
\caption{\label{fig:5} \textbf{Orbital interaction and antibonding analysis of AgTl$_2$I$_3$.} Projected crystal orbital Hamilton population (-pCOHP) and (PDOS) analysis of (a) Ag1-I2/I3, (b) Ag2-I3, and (c) Tl-I1/I2/I3 orbital interactions in AgTl$_2$I$_3$. Positive and negative values of -pCOHP indicate bonding and anti-bonding states, respectively. The energy is shifted to the Fermi level at 0 eV. (d) Schematic of the orbital interaction involving the Tl (6s), Ag (4d), and I (5p) atoms.  (e) Comparison of the maximum antibonding values of -pCOHP and the antibonding range below the Fermi level of AgTl$_2$I$_3$ with those of other thermal insulation materials.}
\end{figure*}

To shed light on the quantum-chemical mechanisms underlying weak bonding in AgTl$_2$I$_3$, we performed the crystal orbital Hamiltonian population (COHP) \cite{j100135a014} calculations based on the DFT wavefunctions to analyze pairwise atomic interactions. This method decomposes the electronic energy of bands into pairwise orbital contributions, enabling the classification of interactions as bonding, antibonding, or nonbonding states. As shown in Figs. \ref{fig:5}a, b, and c, all cation-anion pairs in AgTl$_2$I$_3$ exhibit extended antibonding states below the Fermi level, which arise from the interaction between the 4\textit{d} orbital of Ag, the 6\textit{s} orbital of Tl, and the 5\textit{p} orbital of I. Figure \ref{fig:5}d further illustrates the antibonding states of Ag/Tl-I pairs, visualizing the bonding interaction between Ag (4\textit{d})/Tl (6\textit{s}) and I (5\textit{p}) atoms. These extended antibonding states suggest that atomic coordination weakens the lattice strength by disrupting the interference of quantum mechanical wavefunctions between orbitals from different pairs. The formation of such \textit{s}-\textit{p}/\textit{p}-\textit{d} antibonding states at the top of the valence band is a common feature in many metal chalcogenides and halides, and it is a key characteristic of thermal insulating materials \cite{acharyya2022glassy, adfm202304607, jacs3c05091, zhang2023dynamic}.

We compared the COHP values (including the maximum absolute value and the range of antibonding energy extension) of the strongest antibonding state between Ag(4\textit{d}) and I(5\textit{p}) in AgTl$_2$I$_3$ with those in other materials exhibiting ultralow thermal conductivity, as shown in Fig. \ref{fig:5}e. The maximum COHP value for the Ag(4\textit{d})−I(5\textit{p}) interaction in AgTl$_2$I$_3$ is significantly higher than those in other materials, and the antibonding state of Ag(4\textit{d})−I(5\textit{p}) exhibits a broader energy extension range. The extended strong antibonding interactions between atomic pairs in AgTl$_2$I$_3$ promote the formation of a soft lattice and weak bonding, which lead to low phonon group velocities. This explains the simultaneous suppression of both particle-like phonon propagation and wave-like interband tunneling.

In summary, we synthesized polycrystalline AgTl$_2$I$_3$ with a polyhedron-sharing structure and reported its ultralow $\kappa_{\rm L}$ at room temperature through state-of-the-art experiments and theories. The thermal conductivity is one of the lowest recorded among bulk inorganic crystalline materials. Unlike comparable materials with no further suppression of thermal conductivity above room temperature, the $\kappa$ value decreases to 0.17 $\rm W m^{-1} K^{-1}$ at 523 K. Through crystal structure and chemical bond analysis, we revealed extended antibonding states in the structure, which lead to a softened lattice and weak bonding. These features result in rattling-like thermal vibrations and low phonon group velocities, suppressing both particle-like phonon propagation and wave-like tunneling thermal transports. Our work provides a pathway for exploring materials with ultralow thermal conductivity through the analysis of lattice dynamics and chemical bonding.

\section{Methods}
\paragraph{Sample preparation and characterization.} The polycrystalline AgTl$_2$I$_3$ sample was prepared using stoichiometric amounts of AgI (powder, 99.9\%) and TlI (ball, 99.9\%) precursors. The precursors were carefully weighed and placed into a graphite tube, which was then loaded into an evacuated silica tube and sealed under a vacuum of approximately $10^{-4}$ Pa. The sealed tube was heated to 773 K over a period of five hours, maintained at this temperature for 24 hours, and then gradually cooled to 473 K within six hours, where it was held for an additional 48 hours. Afterward, the furnace was turned off, allowing the sample to cool naturally to room temperature. The obtained ingot was grinded into fine powder and further sintered via spark plasma sintering (SPS) at 523 K for five minutes under a pressure of 90 MPa. The final sintered pellet achieved a density of 96\% of the theoretical density. 

High-resolution PXRD data of the synthesized AgTl$_2$I$_3$ powder were measured at 300 K using a Bruker D8 Advance Vario 1 two-circle diffractometer ($\theta\sim2\theta$, Bragg−Brentano mode). The instrument was equipped with Cu K$\alpha$1 radiation ($\lambda$ = 1.5406 \AA) and a Johansson-type Ge (111) monochromator. Rietveld refinement was conducted using the JANA2020 \cite{Jana2020} crystallographic computing system. The relevant refined results are summarized in Table S1.

\paragraph{Heat capacity and thermal transport measurements.} The thermal conductivity ($\kappa$) was determined using the formula $\kappa = \rho C_p d$, where $\rho$ is the density, $C_p$ is the isobaric heat capacity, and $d$ is the thermal diffusivity. $d$ of the polycrystalline AgTl$_2$I$_3$ sample was measured from 300 to 523 K using a Netzsch LFA 457 laser flash system under a nitrogen atmosphere. For temperatures below 300 K, $\kappa$ and $C_p$ were collected by a Quantum Design Physical Property Measurement System (PPMS). $C_p$ measurements above 300 K were conducted from 300 to 523 K using a Netzsch STA 449F3 system (Fig. S1), while $\rho$ was determined via the Archimedes method.

\paragraph{Sound velocity measurements.} The longitudinal and transverse sound velocities of the sintered AgTl$_2$I$_3$ pellet at 300 K were obtained using the pulse-echo method. A small amount of grease was applied to ensure optimal contact between the sample and the piezoelectric transducers.

\paragraph{Density functional theory calculations.} Density functional theory (DFT) calculations were performed using the Vienna \textit{Ab initio} Simulation Package (VASP) \cite{kresse1996VASP}, employing projector-augmented wave (PAW) pseudopotentials \cite{PRB5017953} and the Perdew-Burke-Ernzerhof revision for solids (PBEsol) \cite{PBEsol2008} within the generalized gradient approximation (GGA) \cite{1996PBE} for the exchange-correlation functional. The valence electron configurations of Tl (5\textit{d}$^{10}$6\textit{s}$^2$6\textit{p}$^1$), Ag (4\textit{d}$^{10}$5\textit{s}$^1$), and I (5\textit{s}$^2$5\textit{p}$^5$) were considered. The initial AgTl$_2$I$_3$ structure was fully relaxed until atomic forces were reduced to below 10$^{-4}$ eV/\AA. The plane-wave energy cutoff was set to 520 eV, and Brillouin zone sampling was performed using a $\rm {\Gamma}$-centered Monkhorst-Pack \textit{k}-point grid of 6 × 6 × 6. The optimized lattice constants of AgTl$_2$I$_3$ were found to be \textit{a} = \textit{b} = 10.329 \AA \ and \textit{c} = 19.421 \AA, which are in good agreement with the experimental values shown in Table S1. To calculate the anharmonicity parameters ($\sigma$) \cite{PRM4083809} and atomic anisotropic displacement parameters, \textit{ab initio} molecular dynamics (AIMD) simulations were carried out with a 2 × 2 × 2 supercell (144 atoms) using a plane-wave cutoff of 400 eV and a $\rm {\Gamma}$-centered 1 × 1 × 1 \textit{k}-point mesh. The energy convergence criterion was set to 10$^{-4}$ eV, with a time step of 2.0 fs and 10000 simulation steps. For static calculations, a 520 eV energy cutoff and a 10$^{-8}$ eV convergence criterion were applied, using a 3 × 3 × 3 $\rm {\Gamma}$-centered \textit{k}-point grid to sample the Brillouin zone of AgTl$_2$I$_3$. 

\paragraph{Extraction of interatomic force constants.} A total of 70 configurations were randomly selected from AIMD simulations at each temperature to form the training set, and atomic forces were then obtained through static DFT calculations. The atomic displacements and forces were used to fit and extract the second-order interatomic force constants (2nd-IFCs) using the self-consistent phonon (SCPH) theory \cite{PRB1572, SCPH2014} implemented in the Hiphive package \cite{eriksson2019hiphive}. The harmonic (second-order) interatomic force constants (IFCs) at 0 K were calculated using the small displacement method implemented in the Phonopy package \cite{TOGO20151} with a displacement of 0.01 \AA. The third- and fourth-order force constants were extracted using the Hiphive package \cite{eriksson2019hiphive}. The harmonic terms at 0 K were subtracted from the force-displacement data before training the cluster space, ensuring that the anharmonic strengths of the third- and fourth-order IFCs were accurately captured. Based on the convergence test (Fig. S2), we determined that the cutoff distances for the second-order, third-order, and fourth-order IFCs are 8.0, 5.5, and 4.0 \AA, respectively.

\paragraph{Anharmonicity parameter calculations.} We adopted the dimensionless anharmonicity parameter $\sigma$ proposed by Knoop et al. \cite{PRM4083809, PRL130236301} to quantitatively evaluate the degree of lattice anharmonicity, defined as
\begin{equation}
\begin{aligned}
\sigma(T) \equiv \frac{\sigma\left[F^{\mathrm{A}}\right]_T}{\sigma[F]_T}=\sqrt{\frac{\sum_{I, \alpha}\left\langle\left(F_{I, \alpha}^{\mathrm{A}}\right)^2\right\rangle_T}{\sum_{I, \alpha}\left\langle\left(F_{I, \alpha}\right)^2\right\rangle_T}}
\end{aligned}
\end{equation}
where $\langle \rangle$ denotes the ensemble average. $F_{I, \alpha}^A$ and $F_{I, \alpha}$ are the anharmonic and total \textit{ab initio} atomic forces of atom $I$ along $\alpha$ direction, and $F_{I, \alpha}^A$ is obtained by subtracting the harmonic force from $F_{I, \alpha}$. This provides a systematic measure of anharmonicity by evaluating the standard deviation of the distribution of anharmonic force components during AIMD simulations.

\paragraph{Lattice thermal conductivity calculations.}  We calculated the three- and four-phonon linewidths using the extended ShengBTE package \cite{HAN2022108179} within the formulae derived by Feng and Ruan et al. \cite{PRB93045202}. The scalebroad was set to 1.0, and a 9 × 9 × 9 \textit{q}-point mesh was adopted to compute the phonon linewidths. Both particle-like phonon propagation $\kappa_{\rm L}^{\rm p}$ and wave-like phonon tunneling $\kappa_{\rm L}^{\rm c}$ contributions within the unified thermal transport theory \cite{UT2019} are calculated using the Phono3py package \cite{PRB91094306}, which allows the inclusion of four-phonon linewidths when combining with our in-house extension. The formula is as follows
\begin{equation}
\begin{aligned}
\kappa_{\rm L}^{\rm p/c} & =\frac{\hbar^2}{N V k_{\mathrm{B}} T^2} \sum_{\boldsymbol{q}} \sum_{s, s^{\prime}} \frac{\omega_{\boldsymbol{q}, s} +\omega_{\boldsymbol{q}, s^{\prime}}}{2} \boldsymbol{u}_{\boldsymbol{q}, s, s^{\prime}} \boldsymbol{u}_{\boldsymbol{q}, s^{\prime}, s} \\
& \times \frac{\omega_{\boldsymbol{q}, s} n_{\boldsymbol{q}, s}\left(n_{\boldsymbol{q}, s} + 1\right) + \omega_{\boldsymbol{q}, s^{\prime}} n_{\boldsymbol{q}, s^{\prime}}\left(n_{\boldsymbol{q}, s^{\prime}} + 1\right)}{4\left(\omega_{\boldsymbol{q}, s}-\omega_{\boldsymbol{q}, s^{\prime}}\right)^2+\left({\rm {\Gamma}}_{\boldsymbol{q}, s} + {\rm {\Gamma}}_{\boldsymbol{q}, s^{\prime}}\right)^2} \\
& \times\left({\rm {\Gamma}}_{\boldsymbol{q}, s} + {\rm {\Gamma}}_{\boldsymbol{q}, s^{\prime}}\right)
\end{aligned}
\end{equation}
where $\hbar$, \textit{N}, $\textit{V}$, $k_B$, $T$, and \textit{n} are the reduced Planck constant, the total number of sampled wave vectors, the volume of the unit cell, the Boltzmann constant, temperature, and phonon occupation probability, respectively. $\boldsymbol{u}_{\boldsymbol{q}, s, s^{\prime}}$, $\omega_{\boldsymbol{q}, s}$, and ${\rm {\Gamma}}_{\boldsymbol{q}, s}$ represent the cross-group velocity matrix, the phonon frequency, and the linewidth of the phonon mode with index \textit{s} for the wave vector $\boldsymbol{q}$, respectively. If $s$ = $s^{\prime}$, the conventional particle-like phonon propagation ($\kappa^{\rm p}_{\rm L}$) is calculated, and when $s \neq s^{\prime}$, we obtain the wave-like interband tunneling ($\kappa^{\rm c}_{\rm L}$). The lattice thermal conductivities of AgTl$_2$I$_3$ in the \textit{x} and \textit{y} directions are equal, and the average value of the three directions is compared with the measured thermal conductivity of polycrystalline AgTl$_2$I$_3$.

\textbf{Acknowledgments}
This work is supported by the Research Grants Council of Hong Kong (C7002-22Y, 17318122, and N\_HKU702/24) and the Guangdong Major Project of Basic and Applied Basic Research (2020B0301030001). X.S. acknowledges funding supported by the Fundamental Research Funds for the Central Universities (D5000250021). The authors are grateful for the research computing facilities offered by the ITS, HKU.

\textbf{Author contributions}
R.C. and Y.C. conceived the idea and designed the project. R.C., C.W., and N.O. wrote the code and performed the calculations. X.S. performed the experiments. R.C., X.S., and Y.C. wrote the paper.

\textbf{Data availability}
The data that support the findings of this study are available on request.

\textbf{Code availability}
The DFT calculations and AIMD simulations were performed using the VASP package \cite{kresse1996VASP, 1999PAW}. The interatomic force constants were extracted using the Hiphive package \cite{eriksson2019hiphive}. Phonon linewidths and thermal conductivity were calculated with the ShengBTE \cite{HAN2022108179} and Phono3py packages \cite{PRB91094306, UT2019}, respectively. Chemical bonding was analyzed using the crystal orbital Hamiltonian population (COHP) method, as implemented in the LOBSTER package \cite{j100135a014, jp202489s}. GPUMD package  \cite{FAN201710} was used to perform machine learning neuroevolution potential (NEP) training and homogeneous nonequilibrium molecular dynamics (HNEMD) simulations.

\textbf{Competing interests}
The authors declare no conflict of interest.

\textbf{Keywords}
anharmonic lattice dynamics, phonon, thermal transport, antibonding states

\bibliography{AgTl2I3}
\end{document}